\documentclass[aps,twocolumn,showpacs,preprintnumbers,amsmath,amssymb,prd]{revtex4-1}

\usepackage{graphicx}
\usepackage{dcolumn}
\usepackage{amssymb}

\newcommand{\be}{\begin{equation}}
\newcommand{\ee}{\end{equation}}
\newcommand{\bea}{\begin{eqnarray}}
\newcommand{\eea}{\end{eqnarray}}

\newcommand{\lcm}{\begin{changemargin}{-1.8cm}{0.25cm}}
\newcommand{\ecm}{\end{changemargin}}
\newcommand{\tbcm}{\begin{changemargin}{-0.8cm}{-0.25cm}}
\newcommand{\tecm}{\end{changemargin}}

\begin{document}
\def\v#1{{\bf #1}}

%\preprint{APS/123-QED}

\title{A Novel Mechanism to Generate FFLO States in Holographic Superconductors}

\author{James Alsup}
\email{jalsup@umflint.edu}
\affiliation{Computer Science, Engineering and
Physics Department, The University of Michigan-Flint, Flint, MI
48502-1907, USA}
\author{Eleftherios Papantonopoulos}
\email{lpapa@central.ntua.gr}
\affiliation{Department of Physics, National Technical University
of Athens, GR-15780 Athens, Greece}
\author{George Siopsis}
\email{siopsis@tennessee.edu}
\affiliation{Department of Physics and Astronomy, The University of
Tennessee, Knoxville, TN 37996-1200, USA}
%\affiliation{{$^{\natural}$}Computer Science, Engineering and
%Physics Department, The University of Michigan-Flint, Flint, MI
%48502-1907, USA\\
%%\email{pkotetes@central.ntua.gr}
%$^{*}$Department of Physics, National Technical University
%of Athens, GR-15780 Athens, Greece\\
%$^{\flat}$Department of Physics and Astronomy, The University of
%Tennessee, Knoxville, TN 37996 - 1200, USA}

\begin{abstract}
We discuss a novel mechanism to set up a gravity dual of FFLO
states in strongly coupled superconductors.  The gravitational
theory utilizes two $U(1)$ gauge fields and a scalar field coupled
to a charged AdS black hole. The first gauge field couples with
the scalar sourcing a charge condensate below a critical
temperature, and the second gauge field provides a coupling to
spin in the boundary theory. The scalar is neutral under the
second gauge field. By turning on an interaction between the
Einstein tensor and the scalar, it is shown that, in the low
temperature limit, an inhomogeneous solution possesses a higher
critical temperature than the homogeneous case, giving rise to
FFLO states.

\end{abstract}

\date{\today}

\pacs{11.25.Tq, 04.70.Bw, 71.45.Lr, 71.27.+a} \maketitle

%11.25.Tq    Gauge/string duality
%04.70.Bw    Classical black holes
%71.45.Lr    Charge-density-wave systems (see also 75.30.Fv Spin-density waves)
%71.27.+a    Strongly correlated electron systems; heavy fermions

The AdS/CFT correspondence, which was discovered in string theory, has opened up a broad avenue for the exploration of condensed matter systems at strong coupling. By using a holographic principle, these systems (described by gauge field theories) are mapped onto weakly coupled gravitational systems of one additional dimension, in which physical quantities can be computed. This holographic principle (gauge theory / gravity duality) has
been applied to the study of conventional and unconventional
superfluids and superconductors \cite{HartnollPRL101}, Fermi
liquids \cite{Bhattacharyya:2008jc}, and quantum phase transitions
\cite{Cubrovic:2009ye}.

The high-Tc superconductors, such as cuprates and iron pnictides,
are examples of unconventional superconductors which   exhibit
competing orders that are related to the breaking of the lattice
symmetries. This breaking introduces inhomogeneities and a study
of the effect of inhomogeneity of the pairing interaction in a
weakly coupled BCS system \cite{Martin:2005fk} suggests that
inhomogeneity might play a role in high-Tc superconductivity. In
an effort to explain this behavior a ``striped" superconductor
was proposed \cite{Berg:2009fk}. Holographic striped
superconductors were discussed in \cite{Flauger:2010tv} where a
modulated chemical potential was introduced and  it was shown that
below a critical temperature superconducting stripes develop.
Properties of the striped superconductors and backreaction
effects were studied in \cite{Hutasoit:2011rd,Ganguli:2012up}.
Striped phases were also found in electrically charged RN-AdS
black branes that involve neutral pseudo-scalars
\cite{Donos:2011bh}.

Inhomogeneous phases also appear when a strong external magnetic
field coupled to the spins of the conduction electrons is applied
to a high-field superconductor. This results in a separation of
the Fermi surfaces corresponding to electrons with opposite spins
(for a review see \cite{Casalbuoni:2003wh}). If the separation is
too high, the pairing is destroyed and there is a transition from
the superconducting state to the normal one (paramagnetic effect).
An intriguing new state of matter at the transition
point was proposed by Fulde and Ferrell \cite{Fulde} and
Larkin and Ovchinnikov \cite{Larkin} (the FFLO state) but it has
not been found experimentally so far. This state is characterized
by a space modulated order parameter, corresponding to an electron
pair having  nonzero total momentum.

A way to understand the formation of  the FFLO phase in a
superconductor-ferromagnetic system (S/F) is to use the
generalized Ginzburg-Landau expansion.  In order to describe the paramagnetic
effect in the presence
of a strong external magnetic field, the usual $|\psi|^4$-Ginzburg-Landau functional has to be
modified with coefficients in the functional which depend also on
the magnetic field. In this case, the $(B,T)$ phase diagram
exhibits a different behavior indicating that  the minimum of the
functional does not correspond to a uniform state, and a spatial
variation of the order parameter decreases the energy of the
system. To describe such a situation, it is necessary to add a
higher-order derivative term in the expansion of the Ginzburg-Landau
functional (for a detailed account see \cite{Buzdin:2005zz}).

There are several studies of the  behavior of holographic superconductors in
the presence of an external magnetic field. Non-trivial spatially
dependent solutions have been found, like the droplet
\cite{Albash:2008eh} and vortex solutions with integer winding
number \cite{Albash:2009iq, Montull:2009fe, Maeda:2009vf}. An
analytic study on holographic superconductors in an external
magnetic field was carried out in \cite{Ge:2010aa}. In a model
resulting from  a consistent truncation of type IIB string theory,
 anisotropic solutions at low temperature were found
\cite{ABKProchazka}, showing similarity between the phase diagrams
of holographic superfluid flows and those of ordinary
superconductors with an imbalanced chemical potential. A holographic
superconducting model with unbalanced fermi mixtures at strong
coupling was discussed in \cite{Bigazzi:2011ak}.  The charge and
spin transport properties of the model were studied, but the phase
diagram did not reveal the  occurrence of FFLO-like inhomogeneous
superconducting phases.

In our recent work \cite{Alsup:2012ap}, we proposed a gravity dual
of FFLO states in strongly coupled superconductors. The gravity
sector consisted of two $U(1)$ gauge fields and a scalar
field. The first gauge field had a non-zero scalar potential term
which was the source of the charge condensate in the boundary theory through its coupling to the scalar
field. The second $U(1)$
gauge field corresponded to an effective magnetic field acting on
the spins in the boundary theory. The scalar field was neutral
under the second $U(1)$ gauge field. We looked first at the
behavior of the system at or above the critical temperature. The
Eintein-Maxwell system was solved by a dyonic black hole with
electric and magnetic charges,
as in \cite{Bigazzi:2011ak}. At the critical
temperature,
 the system underwent a second-order phase transition
and the black hole acquired hair. To find the critical temperature,
we worked in the grand canonical ensemble and solved the scalar equation in
the background of the dyonic black hole.   It was found that the
system possessed inhomogeneous solutions for the scalar field,
which however always gave a transition temperature lower than the
maximum transition temperature (i.e. critical temperature) of the homogeneous solution. Therefore the homogeneous solution was always dominant.

Next, we turned on an interaction term of the magnetic field to the
scalar field of the generalized Ginzburg-Landau gradient type (in
a covariant form). The scalar field equation was modified and the
resulting inhomogeneous solutions gave a transition temperature
which was higher than the one of the homogeneous solutions. We attributed this
behavior of the system to the appearance of FFLO states. We noted
that the appearance of the FFLO states was more pronounced as
$T_c/\mu \rightarrow 0$, and the magnetic field of the second
$U(1)$ gauge group was large.

In this letter, we propose a novel mechanism for the generation
of the gravity dual of FFLO states in the low temperature limit. In
our previous work we showed that, in order to generate the FFLO phase, we
needed a direct coupling of the magnetic field to the scalar field.
We will show that this interaction term can be effectively generated
through the coupling of the Einstein tensor to the
scalar field. The reason is that since the electromagnetic fields
backreact on the metric, the Einstein tensor has encoded the
information of these fields.

As before, the bulk theory consists of two $U(1)$ gauge
fields and a scalar field. The first gauge field has a non-zero
scalar potential term and the second $U(1)$ gauge field corresponds to a
chemical potential (imbalance) for spin.
The scalar field is neutral under the second $U(1)$ gauge field.
Note, too, the second $U(1)$
 is self-dual under $\vec E \leftrightarrow \vec B$ and alternatively the boundary
 theory can be understood in terms of a magnetic field instead of the chosen spin
 chemical potential.

The interaction between the Einstein
tensor and the scalar field is most often
seen in scalar-tensor theories. The interest stems from the
galilean symmetry  of the system where the action is invariant
under shifts of field derivatives by a constant vector. Thereby
the higher-derivative theory has only second order equations of
motion \cite{galilean}. It was shown that this term acts as an
effective cosmological constant and produces an early entrance into a
quasi-de Sitter stage as well as a smooth exit \cite{Sushkov}.  Cosmic evolution for vanishing cosmological constant has also been
investigated in \cite{STcosmo}.
  The coupling has  been realized in string
cosmology from an effective heterotic action, up to $\alpha'$
corrections \cite{MaedaOhtaWakebe} and also in N = 1
four-dimensional new-minimal supergravity theories
\cite{Farakos:2012je}.

Moreover, interest away from cosmology has developed as the
interaction has been used to study phase transitions for vanishing
cosmological constant
 \cite{Kolyvaris} and effects
on conventional holographic superconductors employing anti-de
Sitter space \cite{Chen:2010hi}. The presence of this term
modifies the scalar field equation  and the resulting
inhomogeneous solutions give a transition temperature which is
higher than the homogeneous solutions. We attribute this behavior to FFLO states. Note that as
before, the appearance of the FFLO states is more pronounced as
$T_c/\mu \rightarrow 0$ and the gauge field of the second
$U(1)$ gauge group is near its maximum value.

Consider the action \bea S &=& \nonumber \\ & \int & d^4 x
\sqrt{-g} \left[ \frac{R + 6/L^2}{16\pi G} - \frac{1}{4} F_{AB}
F^{AB}  - \frac{1}{4} \mathcal{F}_{AB} \mathcal{F}^{AB}\right],
\nonumber
\\ \eea where $F_{AB} =
\partial_A A_B - \partial_B A_A$, $\mathcal{F}_{AB} =
\partial_A\mathcal{A}_B -\partial_B \mathcal{A}_A$ are the field
strengths of the $U(1)$ potentials $A_A$ and $\mathcal{A}_A$,
respectively. We set $L=8\pi G = 1$.

The Einstein-Maxwell equations,
\bea
R_{\mu\nu} &-&\frac{1}{2} g_{\mu\nu} R-\frac{3}{L^2}g_{\mu\nu}=\nonumber\\
&&\frac{1}{2} \big[F_{\mu\sigma}F^{\sigma}_\nu-\frac{1}{4} g_{\mu\nu} F^2+\mathcal F_{\mu\sigma}\mathcal F^{\sigma}_\nu-\frac{1}{4} g_{\mu\nu} \mathcal{F}^2\big],\nonumber\\
\nabla_\mu &F&^{\mu\nu} = 0,\nonumber\\
\nabla_\mu &\mathcal{F}&^{\mu\nu} = 0,
\eea
admit a solution which is a
four-dimensional AdS black hole of two $U(1)$ charges, \be ds^2 =
\frac{1}{z^2} \left[ - h(z) dt^2 + \frac{dz^2}{h(z)} + dx^2 + dy^2
\right],
\label{eqmetric1}
\ee with the horizon radius set at
$z=1$.

The two sets of Maxwell equations admit solutions of the form,
respectively, \be\label{eq3} A_{t} = \mu \left( 1 - z \right) \ , \ \ A_z
=A_x=A_y =0, \ee and \be\label{eq4} \mathcal{A}_t = \delta\mu (1-z) \ , \ \
\mathcal{A}_z = \mathcal{A}_x = \mathcal{A}_y = 0, \ee with
corresponding field strengths having non-vanishing components for
electric fields in the $z$-direction, respectively,
\be\label{eq5} F_{tz} = - F_{zt} = \mu \ \ , \ \ \ \ \mathcal{F}_{tz} =
-\mathcal{F}_{zt} = \delta\mu~. \ee Then from the Einstein
equations we obtain \be\label{eq6}  h(z) = 1 -\left( 1 +
\frac{\mu^2+\delta\mu^2}{4} \right) z^3 + \frac{\mu^2 +\delta\mu^2}{4} z^4 ~.\ee
The Hawking temperature is \be\label{eq7} T = - \frac{h'(1)}{4\pi}
= \frac{3}{4\pi } \left[ 1 - \frac{\mu^2+\delta\mu^2}{12}
\right]. \ee In the limit $\mu, \delta\mu \to 0$ we recover the
Schwarzschild black hole.

Next, we consider a scalar field $\phi$, of mass $m$, and $U(1)^2$
charge $(q,0)$, coupled to the Einstein tensor. The action is
\bea\label{eq8} S = &-&\int d^4 x \sqrt{-g} \Big[  g^{AB}  (D_A \phi)^\ast D_B\phi + m^2 (1-3\xi) |\phi|^2 \nonumber\\
&-& \xi
G^{AB} (D_A \phi)^\ast D_B\phi \Big],\eea
where $D_A = \partial_A + iqA_A$ and $G_{AB}$  is the Einstein
tensor. $\xi$ is the new coupling constant determining the strength of the interaction between the scalar field and the Einstein tensor.
We also included a convenient $\xi$-dependent factor in the mass term. We shall consider the range of $\xi$ for which the factor is positive,
\be\label{eq10xi} \xi < \frac{1}{3} ~.\ee
Firstly, we consider the conventional case setting $\xi =0$.
The asymptotic
behavior (as $z\to 0$) of the scalar field is \be\label{eq9} \phi \sim z^\Delta \ \ , \ \ \ \ \Delta (\Delta -3) = m^2~. \ee
For $m^2 \geq -5/4$, there is only one normalizable mode.  However, for the range, $-9/4 \leq m^2 <-5/4$, there are two allowable choices of $\Delta$,
\be\label{eq10} \Delta = \Delta_\pm = \frac{3}{2} \pm \sqrt{\frac{9}{4} +m^2}~, \ee
leading to two distinct physical systems.  

As we lower the temperature, an instability arises and the system undergoes a second-order phase transition with the black hole developing hair.
This occurs at a critical temparture $T_c$ which is found by solving the scalar wave equation in the above background,
\be\label{eq19} \partial_z^2 \phi + \left[ \frac{h'}{h} - \frac{2}{z}
\right] \partial_z \phi + \frac{1}{h} \nabla_2^2\phi - \frac{1}{h}
\left[ \frac{m^2}{z^2} - q^2 \frac{A_t^2}{h}  \right] \phi = 0, \ee
with the metric function $h$ given in \eqref{eq6} and the electrostatic potential $A_t$ in \eqref{eq3}.

Although the wave equation \eqref{eq19} possesses $(x,y)$-dependent solutions, the symmetric solution dominates and the hair that forms has no $(x,y)$ dependence. To see this, let us introduce $(x,y)$-dependence and consider a static scalar field which is an eigenstate of the two-dimensional Laplacian,
\be\label{eq11}  \nabla_2^2\phi = -\tau \phi \ \ , \ \ \ \ \tau > 0 ~.
\ee
For example, if $\phi$ varies sinusoidally in the $x$-direction, $\phi \sim e^{iQx}$, then $\tau = Q^2$ and the modulation is realized in the boundary CFT through the order parameter $\langle \mathcal O \rangle \sim e^{iQx}$.  It is also possible for $\phi$ to be rotationally symmetric in the $(x,y)$ plane, $\phi \sim J_0(\sqrt{\tau(x^2+y^2)})$.  For $\tau=0$, we recover the homogeneous solution.

Upon factoring out the $(x,y)$ dependence,
\be \phi = Y(x,y) \psi (z)~, \ee
where $Y(x,y)$ is an eigenfunction of the two-dimensional Laplacian with eigenvalue $-\tau$ (eq.\ \eqref{eq11}), the scalar field is represented by $\psi(z)$ and the wave equation becomes
\be\label{eqw} \psi''
+ \left[ \frac{h'}{h} - \frac{2}{z} \right] \psi' - \frac{\tau}{h}
\psi - \frac{1}{h} \left[ \frac{m^2}{z^2} - q^2 \frac{A_t^2}{h}
\right] \psi = 0. \ee
Before we proceed with a discussion of solutions, notice that there is a scaling
symmetry \bea  &z& \ \to \lambda z \ , \ \ x\to \lambda x \ , \ \
\tau \to \tau /   \lambda^2 \ , \nonumber \\  \ \ &\mu& \ \to \mu
/\lambda \ , \ \ \delta\mu \to \delta\mu /\lambda \ , \ \ T
\to T/\lambda~. \eea
This means that the system possesses a scale which we have fixed for simplicity of notation. This arbitrary scale is often taken to be the radius of the horizon $r_+$, after changing coordinates to
\be z =\frac{r_+}{r}~. \ee
Since we fixed the scale, we should only be reporting on
scale-invariant quantities, such as $T/\mu$, $\delta\mu / \mu$,
$\tau/\mu^2$, etc.
It is also convenient to introduce the scale-invariant parameter
\be \beta = \frac{\delta\mu}{\mu} \ee
to describe the effect of the chemical potential imbalance.

We shall be working in the grand canonical ensemble at fixed chemical potentials $\mu$ and $\delta\mu$ (or $\beta$).
The ensemble is defined uniquely by specifying the parameters $q$ and $\Delta$. One can then vary $\tau$, which parametrizes the solutions, to study the behavior of the system.

Since we fixed the scale, we shall solve the wave equation \eqref{eqw} for fixed values of
the scale-invariant parameters $\tau/\mu^2$ and $\beta$, while demanding regularity at the horizon and $\psi \sim z^{\Delta_+}$ at the boundary. Thus, we obtain $\mu = \mu_0$ as an eigenvalue. More precisely, we obtain $\mu/r_+$ as an eigenvalue, where $r_+$ is the scale in the system. Since the chemical potential $\mu$ is fixed (grand canonical ensemble), the solution of the wave equation \eqref{eqw} fixes the scale ($r_+ = r_{+0}$, so that $\mu_0 \to \mu /r_{+0}$) , and therefore the transition temperature $T_0$ below which a mode with the given $\tau$ may develop. We obtain
\be\label{eqT} \frac{T_0}{\mu_{\mathrm{eff}}} =  \frac{3}{4\pi \mu_{\mathrm{eff,0}}} \left[ 1 - \frac{\mu_{\mathrm{eff,0}}^2}{12}
\right]~, \ee
which is of the same form as a Reissner-Nordstr\"om black hole with effective chemical potential
\be \mu_{\mathrm{eff}}^2 = \mu^2 (1+\beta^2)~. \ee
The maximum transition temperature is the \emph{critical temperature} $T_c$ of the system. As we cool down the system in its normal state, the transition temperature $T_0=T_c$ is reached first and the mode with the corresponding $\tau$ is the first to develop.

In the homogeneous case, $\tau = 0$, the maximum transition temperature is obtained for $\beta =0$. In this case, we recover the Reissner-Nordstr\"om black hole. As we increase $\beta$, the temperature \eqref{eqT} decreases.
% For a given $\beta>0$, the black hole is of the Reissner-Nordstr\"om form with effective chemical potential
The scalar wave equation is the same as its counterpart in a Reissner-Nordstr\"om background, but with effective charge
\be q_{\mathrm{eff}}^2 = \frac{q^2}{1+\beta^2}~, \ee
so that $q_{\mathrm{eff}} \mu_{\mathrm{eff}} = q\mu$.

It is known \cite{HartnollPRL101} that the instability \cite{Breitenlohner:1982jf} occurs for all values of $q_{\mathrm{eff}}$, including $q_{\mathrm{eff}}=0$, if $\Delta \le \Delta_\ast$, where $\Delta_\ast =\Delta_+$ for $m^2 = - \frac{3}{2}$, or explicitly,
\be \Delta_\ast = \frac{3+\sqrt{3}}{2} \approx 2.366 ~. \ee
For $\Delta \le \Delta_\ast$, $\beta$ can increase indefinitely. The transition temperature $T_0$ for $\tau =0$ has a minimum value as a function of $\beta$, and as $\beta\to\infty$, $T_0$ diverges.

For $\Delta > \Delta_\ast$, $q_{\mathrm{eff}}$ has a minimum value at which the transition temperature vanishes and the black hole attains extremality. This is found by considering the limit of the near horizon region \cite{Horowitz:2009ij,AST}. One obtains
\be\label{eqqmin1} q_{\mathrm{eff}} \ge q_{\mathrm{min}} \ \ , \ \ \ \ q_{\mathrm{min}}^2 = \frac{3+2\Delta (\Delta -3)}{4}. \ee
At the minimum ($T_0=0$), $\mu_{\mathrm{eff}}^2 =12$, and $\beta$ attains its maximum value,
\be\label{eqbmax} \beta \le \beta_{\mathrm{max}} \ \ , \ \ \ \ \beta_{\mathrm{max}}^2 = \frac{q^2}{  q_{\mathrm{min}}^2} -1~. \ee
This limit is
reminiscent of the Chandrasekhar and Clogston limit
\cite{instability} in a S/F system, in which a ferromagnet at
$T=0$ cannot remain a superconductor with a uniform condensate.

In the inhomogeneous case ($\tau\ne 0$), the above argument still holds with the replacement $m^2 \to m^2 + \tau$.
The effect of this modification is to increase the minimum effective charge to
\be\label{eqqmin} q_{\mathrm{min}}^2 = \frac{3+2\Delta (\Delta -3)+2\tau}{4}, \ee
and thus decrease the maximum value of $\beta$ \eqref{eqbmax}. We always obtain a transition temperature which is lower than the corresponding transition temperature (for same $\beta$) in the homogeneous case ($\tau=0$). It follows that the critical temperature is the transition temperature of the homogeneous mode, and the latter dominates the condensate.

\begin{figure}[t]
\includegraphics[width=8.3cm]{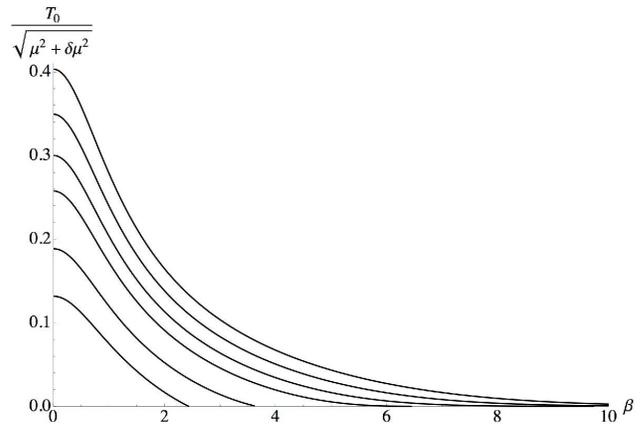}
\caption{The transition temperature for various modes $vs.$ $\beta$ numerically calculated with $q=10$, $\Delta=5/2$, and $\xi = 0$. Starting from the top, on the vertical axis, the lines are $\frac{\tau}{(q\mu)^2} = 0$ (the critical temperature of the system), $.05,~.10,~.15,~.25,$ and $.35$.}
\label{xi0coupling}
\end{figure}

Notice also that $\tau \le \tau_{\mathrm{max}}$, where the maximum value is attained when $q_{\mathrm{min}} =q$ (so that $\beta_{\mathrm{max}} =0$). We deduce from \eqref{eqqmin},
\be\label{eqQmax} \tau_{\mathrm{max}} = 2q^2 - \frac{3}{2} - \Delta (\Delta -3)~. \ee
The supporting numerical results are shown in figure \ref{xi0coupling}.

Now let us consider the effect of coupling to the Einstein tensor by setting $\xi\ne 0$.
The wave equation is modified to
\bea\label{modWave} \psi'' + \left[ \frac{h'}{h} + \frac{f_+'}{f_+} - \frac{2}{z} \right] \psi'
- \frac{\tau}{h}\, \frac{f_-}{f_+} \psi & & \nonumber\\
- \frac{1}{h} \left[ \frac{m^2(1-3\xi)}{z^2f_+} - q^2 \frac{A_t^2}{h} \right] \psi &=& 0~,
\eea
where
\be\label{eqfpm} f_\pm = 1 -3\xi \pm \xi \frac{ \mu_{\mathrm{eff}}^2 }{4} z^4~. \ee
Note that the boundary behavior is unaltered from \eqref{eq10}.

The coupling to the Einstein tensor alters the near horizon limit of the theory so that
\be\label{meff} m^2 \to \frac{m^2(1-3\xi) + \tau f_-(1)}{f_+(1)} ~,\ee
For $\xi < 1/6$, it is easily seen from \eqref{eqfpm} that $f_\pm(1) > 0$, so that the effective mass increases with $\tau$, as in the conventional $\xi =0$ case.

The minimum effective charge is found from the near-horizon geometry in the zero temperature (extremal) limit. Using $\mu_{\mathrm{eff}}^2 = 12$ and $f_\pm (1) = 1 -3\xi\pm 3 \xi $, we obtain
\be\label{eqqmin2} q_{\mathrm{min}}^2 =\frac{3+2(1-3\xi)\Delta(\Delta -3) +2 (1-6\xi)\tau}{4} ~, \ee
to be compared with \eqref{eqqmin}. 

Finally, $\tau$ has a maximum value found by setting $q_{\mathrm{min}} = q$ in \eqref{eqqmin2},
\be\label{eqQmax2} \tau_{\mathrm{max}} = \frac{2 q^2 - \frac{3}{2} - (1-3\xi)\Delta (\Delta -3)}{1-6\xi}~. \ee
Thus, for $\xi < 1/6$, even though the results differ numerically from the case $\xi =0$, they are not qualitatively different. The maximum transition temperature (i.e., the critical temperature of the system) is \emph{always} attained for $\tau =0$ (homogeneous case).  As $\xi$ approaches the critical value $\frac{1}{6}$, the maximum value of $\tau$ diverges.

As we increase $\xi$ past the critical value, i.e., for
\be\label{xibound}
\xi > 1/6~,
\ee
(with $\xi$ still satisfying \eqref{eq10xi}),
the range of $\tau$ extends to infinity ($\tau_{\mathrm{max}}$ is infinite), and the behavior of the system changes \emph{qualitatively}.
For $\xi$ above the bound \eqref{xibound},
the minimum charge \emph{decreases} for $\tau>0$, and therefore the maximum value of $\beta$ \eqref{eqbmax} \emph{increases} compared to the value in the homogeneous case ($\tau =0$). Thus, there is a neighborhood near zero temperature in which the inhomogeneous solution has higher transition temperature than the homogeneous one.
As we increase $\tau$, the corresponding transition temperature increases. This expected behavior is also seen numerically.

\begin{figure}[t]
\includegraphics[width=8.5cm]{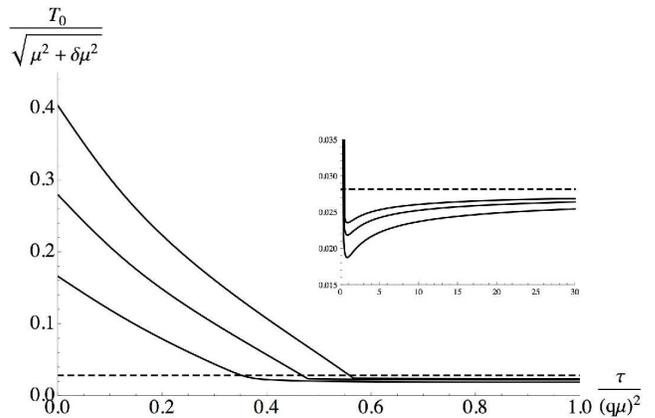}
\caption{The top of the graph corresponds to $\beta=0$, with lines $\beta=1, 2$ below for all values of $\tau$.  The transition temperature of the homogeneous solution ($\tau=0$) is found to be the largest for this range of $\beta$.  We used $q=10$, $\Delta = 5/2$, and $\xi = .20$.  The dotted line represents the asymptotic value for the transition temperature \eqref{asympT}.  The inset is an enlarged view of the lines as they approach the asymptotic (critical) temperature. }
\label{fig2}
\end{figure}

As we keep increasing $\tau$, we are no longer in the zero temperature limit and geometrical considerations near the horizon are no longer applicable. Thus, although the effective mass \eqref{meff} keeps decreasing below the AdS$_2$ BF bound, the latter is no longer relevant, and the wave equation possesses acceptable solutions for all $\tau$. Although we can no longer argue analytically, we analyzed the behavior of the system numerically. As $\tau$ increases, the corresponding transition temperature keeps increasing. The maximum transition temperature, which would be identified with the critical temperature of the system, is attained asymptotically as $\tau\to\infty$ (recall that there is no maximum value of $\tau$ for $\xi > 1/6$).

The value of the critical temperature is found by analytically solving the wave equation in the limit $\tau\to\infty$. It is easy to see by considering an expansion around the horizon that we ought to have
 $f_- = 0$. We deduce
 \be
\mu_{\mathrm{eff,c}}^2  = \frac{4(1-3\xi)}{\xi}~,
\ee
thus determining the asymptotic transition (critical) temperature to be 
 \be\label{asympT}
\lim_{\tau/\mu^2 \to \infty} \frac{T_c}{\mu_{\mathrm{eff}} } \to \frac{3}{4 \pi } \sqrt{\frac{\xi}{1-3\xi} } \left(1-\frac{1}{6\xi}\right),
 \ee
which is dependent solely upon our coupling constant $\xi$.

\begin{figure}[ht!]
\includegraphics[width=8.5cm]{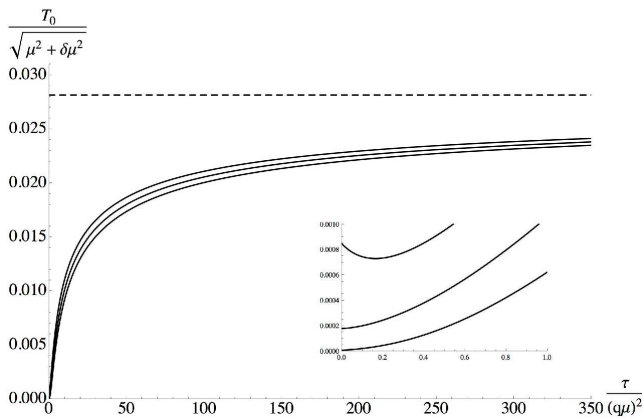}
\caption{The lines, from top to bottom represent $\beta=11.5, 12.5, 13.5$ with  $q=10$, $\Delta = 5/2$, and $\xi = .20$.  The transition temperature of the homogeneous solution ($\tau=0$) is less than that of $\tau/\mu^2 \to \infty$.  The dotted line represents the asymptotic value for the temperature \eqref{asympT} (critical temperature). The inset shows the curves for small $\tau$.}
\label{fig3}
\end{figure}

To find the critical temperature numerically (and confirm the analytic prediction \eqref{asympT}), we fix the chemical potentials $\mu$ and $\delta\mu$ (or $\beta$) and numerically solve the wave equation \eqref{modWave} for all allowed values of $\tau$.  Figure \ref{fig2} displays the transition temperatures of various modes for small values $\beta$.  The plots attain their maximum at the homogeneous mode, $\tau=0$, and therefore the homogeneous solution is dominant.  The low temperature region is probed with larger values of $\beta$.  Our results are plotted in figure \ref{fig3}.  The homogeneous solution possesses a transition temperature that is below the majority of non-zero $\tau$ and hence the inhomogeneous solutions dominate. 
As we increase $\tau$, the corresponding transition temperature increases and approaches the asymptotic value \eqref{asympT}, as expected. The asymptotic value (critical temperature) is an upper bound for the transition temperatures of the various modes.

\begin{figure}[ht!]
\includegraphics[width=8.3cm]{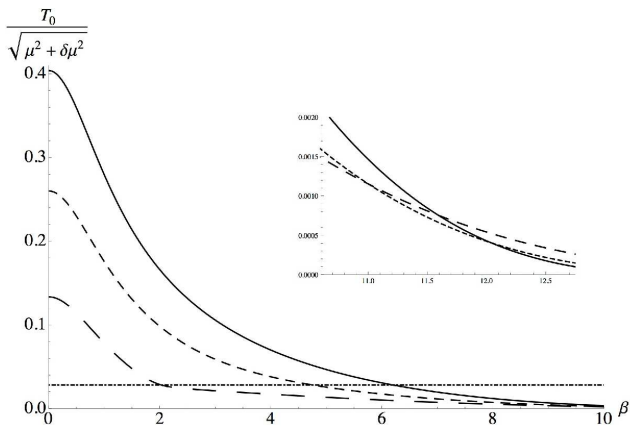}
\caption{The transition temperature of various modes $vs.$ $\beta$ numerically calculated with $q=10$ and $\Delta=5/2$, and $\xi=.20$.  Starting from the top, on the vertical axis, the lines are $\frac{\tau}{(q\mu)^2} = 0,~.15,$ and $.35$, followed by the dash-dotted line representing the asymptotic value for the transition temperature. The crossing between the finite values $\tau$ is shown in the inset.}
\label{fig4}
\end{figure}

Figure \ref{fig4} displays the transition temperature numerically calculated for $\xi=.2$ for select values of $\tau$.  The point where the inhomogeneous solution becomes dominant is found at the crossing between $\tau=0$ and the large $\tau$ asymptotic (critical) temperature seen in the body of the figure.

\begin{figure}[t]
\includegraphics[width=8.3cm]{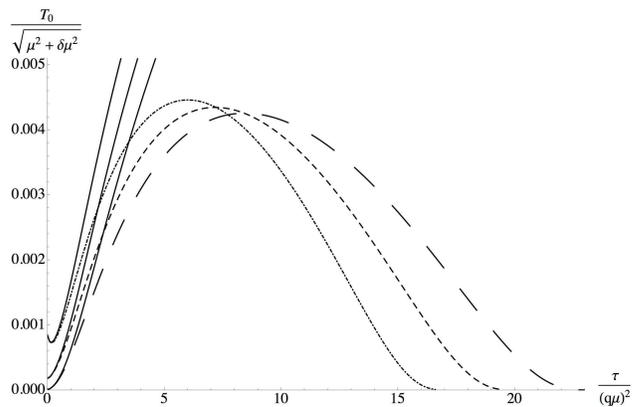}
\caption{The lines, from top to bottom on the left side represent transition temperature for various modes with  $\beta=11.5, 12.5, 13.5$ with  $q=10$, $\Delta = 5/2$, $\xi = .20$.  The solid lines correspond to the $\alpha=0$ solutions while the dashed lines correspond to the cutoff solution with $\alpha=.0001$.  The dot-dashed line is for $\beta=11.5$, the short dashes are used for $\beta=12.5$, and long dashes for $\beta=13.5$.}
\label{fig5}
\end{figure}

In a  condensed matter system with an order parameter possessing wavenumber $Q$, the lattice spacing $a$ is related by $Q \sim 1/a$ \cite{Casalbuoni:2003wh}. In our system, effectively $a\to 0$, which corresponds to $\tau = Q^2 \to \infty$. Therefore, we expect the critical temperature to correspond to $\tau\to\infty$. It would be desirable to include lattice effects so that the critical temperature corresponds to a large but finite value of $\tau$ \cite{latticeDrude,latticeFermion}. An effective way of accomplishing this is by including higher order terms in the Lagrangian. Let us introduce a cutoff that suppresses large momentum ($\tau$) modes in the Einstein coupling term. To do this covariantly, introduce the derivative operator
\be \mathcal{D}^A = \frac{1}{2}\epsilon^{ABCD} \mathcal{F}_{BC} D_D~, \ee
where $\mathcal{F}$ is the field strength of the second $U(1)$ potential, and $D_A$ is the gauge derivative (see eq.\ \eqref{eq8}).
Then modify the action for the scalar field \eqref{eq8} to
\bea\label{eq8m} S = &-&\int d^4 x \sqrt{-g} \Big[  g^{AB}  (D_A \phi)^\ast D_B\phi + m^2 (1-3\xi) |\phi|^2 \nonumber\\
&-& \frac{\xi}{2}
G^{AB} (D_A \phi)^\ast D_B \mathcal{S} \left( -\alpha\mathcal{D}_A \mathcal{D}^A \right) \phi + \mathrm{c.c.} \Big].\eea
The function $\mathcal{S}(x)$ is chosen so that $\mathcal{S} (0) =1$ and $\mathcal{S} (x) \to 0$, as $x\to\infty$. We also introduced a new (small) parameter $\alpha$. It is convenient to choose
\be \mathcal{S} (x) = e^{-x} ~. \ee
The wave equation \eqref{modWave} is modified to
\bea\label{mod2Wave} \psi'' + \left[ \frac{h'}{h} + \frac{f_+'}{f_+} - \frac{2}{z} \right] \psi'
- \frac{\tau}{h}\, \frac{f_-}{f_+} \psi & & \nonumber\\
- \frac{1}{h} \left[ \frac{m^2(1-3\xi)}{z^2f_+} - q^2 \frac{A_t^2}{h} \right] \psi & & \nonumber\\
- 3\alpha\beta^2\mu^2 \tau z^5 \left[ \left( \frac{h'}{h} + \frac{3}{z} \right) \left( 1 - \frac{1}{f_+} \right) + \frac{f_+'}{f_+} \right]\psi &=& 0~,\ \ \ \
\eea
and the functions $f_\pm$ (eq.\ \eqref{eqfpm}) are modified to
\be\label{eqfpmcut} f_\pm = 1 +\xi\left[ -3 \pm \frac{ \mu_{\mathrm{eff}}^2 }{4} z^4 \right] e^{-\alpha \beta^2 \mu^2 \tau z^6}~. \ee
Notice that in the limit $\tau\to\infty$, we have $f_\pm\to 1$, so for large $\tau$, the solutions approach those in the standard case $\xi =0$, in which there is a maximum allowed value of $\tau$ (eq.\ \eqref{eqQmax}).  The supporting numerics are shown in figure \ref{fig5}, where a maximum transition temperature (critical temperature) at finite $\tau$ may be clearly seen.

\emph{In conclusion}, we have developed a gravitational dual
theory for the FFLO state of condensed matter.  The gravitational
theory consists of two $U(1)$ gauge fields and a scalar coupled to
a charged AdS black hole. The first gauge field
produces the instability for a condensate to form, while the
second controls chemical potential associated with spin. In the absence of an
interaction of the Einstein tensor with the scalar field, the system
possesses dominant homogeneous solutions for all allowed values of the spin chemical potential. In the presence of
the interaction term, at low temperatures, the system is shown
to possess a critical temperature for a transition to a scalar field with
spatial modulation as opposed to the homogeneous solution.

It is desirable to fully understand the interplay between the different modes once below the critical temperature.  This will require a non-linear analysis of the Einstein-Maxwell-scalar equations.  Additionally, the dependence of the critical temperature on the modulation wavenumber, which is intertwined with the presence of a lattice is an intriguing aspect. Work in these directions is in progress.
%\vspace{1.2cm}

%\vspace{-.8cm}
%\emph{Dedication.}

\acknowledgments
\vspace{-.2cm}
  J.\ A.\ acknowledges support from the Office of Research at
  the University of Michigan-Flint. G.\ S.\ is supported by the US Department
of Energy under Grant No.\ DE-FG05-91ER40627.

\end{document}